\documentclass[preprint,3p,times]{elsarticle}

\usepackage{lineno,hyperref}
\usepackage[T1]{fontenc}
\usepackage{comment}
\usepackage{amsmath}
\usepackage{amssymb}
\usepackage{graphicx}
\usepackage{epsfig}
\usepackage{latexsym}
\usepackage{pstricks}
\usepackage{pst-plot}
\usepackage{amsmath, amsthm, amsfonts, amssymb}
\usepackage{appendix}
\usepackage[bitstream-charter]{mathdesign}
\modulolinenumbers[5]

\journal{Journal of \LaTeX\ Templates}

\newtheorem{theorem}{Theorem}[section]
\newtheorem{lemma}[theorem]{Lemma}

\begin{document}

\begin{frontmatter}
\title{Lyapunov exponents and Hamiltonian poles in a non Hermitian dynamics}

\author[unlp]{Ignacio S. Gomez\corref{cor1}}
\ead{nachosky@fisica.unlp.ar}

\cortext[cor1]{Corresponding author}
\address[unlp]{Instituto de F\'{i}sica La Plata (IFLP), CONICET, and Departamento de F\'{i}sica, Facultad de Ciencias Exactas, Universidad Nacional de La
Plata, C.C. 67, 1900 La Plata, Argentina}

\begin{abstract}
By means of expressing volumes in phase space in terms of traces of quantum operators, a relationship between the Hamiltonian poles and the Lyapunov exponents in a non Hermitian quantum dynamics, is presented. We illustrate the formalism by characterizing the behavior of the Gamow model whose dissipative decay time, measured by its decoherence time, is found to be inversely proportional to the Lyapunov exponents of the unstable periodic orbits. The results are in agreement with those obtained by means of the semiclassical periodic--orbit approach in quantum resonances theory but using a simpler mathematics.
\end{abstract}


\end{frontmatter}

\nolinenumbers

\section{Introduction}

The interest in the study of non Hermitian Hamiltonians
is related with the interpretation of phenomena
such as nuclear
resonances, dissipation, relaxation of nonequilibrium states, typical of open systems.
In scattering systems one can consider quantum resonances, called ``quasi-stationary states", instead of
scattering solutions \cite{mah69,stockmann,akk94,lan57,lew91}. These quasi-stationary states play in open systems
a similar role as the eigenstates of
closed systems and their eigenvalues are complex
numbers with non zero imaginary part.
Any measurement on a open system drastically
changes its properties by converting discrete energy
levels into decaying quasi-stationary states which can be described by a
non Hermitian Hamiltonian \cite{sieber,kuhl,rotten,moiseyev}. In this context, the characteristic decay times are given by the imaginary part of the complex eigenvalues, i.e. the so called \emph{Hamiltonian poles}. These arise as a result of the analytic extension of a Hamiltonian whose degeneration makes the perturbation theory inapplicable \cite{taylor,bohm,rosenfeld,PRE,antoniou,Gadella,letterpolos,Ordonito-dec,4A}.
Furthermore, non Hermitian Hamiltonians allow to
describe the non-unitary
time evolutions that appear in open
quantum systems \cite{moiseyev}.
Properties of open quantum systems like nonequilibrium phenomena and dissipation can be characterized by the positivity
of the Kolmogorov--Sinai entropy which, in turn, is equal to the sum of all positive Lyapunov exponents due to the Pesin theorem \cite{lich,pesin,young,gutz}.
The characteristic time of these kind of processes is given by the Kolmogorov--Sinai time
which provides a decay time in the phase space as a function of the Lyapunov exponents \cite{krylov1,krylov2,krylov3}.
In addition, in chaotic open quantum systems the Lyapunov exponents and the escape rates of classical trajectories have been characterized by means of semiclassical techniques \cite{semi1,semi2,semi3,semi4}, and also from the strategy of ranking chaos looking at the decay of correlations between states and observables \cite{qsdt,gamowpesin}.

Using the idea of expressing classical quantities in terms of traces of quantum operators as in \cite{qsdt,gamowpesin,qeh0,qeh1,qeh2}, we present a relationship between the Hamiltonian poles and the Lyapunov exponents in a non Hermitian quantum dynamics where the Kolmogorov--Sinai time expresses the contractions and expansions of volumes
in the phase space along their dynamics. The paper is organized as follows. In section 2 we give the preliminaries and mathematical formalism. In section 3 we express the Lyapunov exponents in terms of the Hamiltonian poles by means of the non-unitary evolution of a little volume element in phase space. In section 4 we illustrate the formalism by applying it to the Gamow model. In section 5 we discuss the results with regard the quantum resonances theory.
In section 6 some conclusions and future research directions are outlined.


\section{Preliminaries}

\subsection{Kolmogorov--Sinai time and Pesin theorem}

The characteristic time for a nonequilibrium process
in a mixing dynamics is the Kolmogorov--Sinai time (KS--time) $\tau_{KS}$, which measures the necessary time to take a number of initially close phase points
to uniformly distribute over the energy surface. Moreover, $\tau_{KS}$ is inversely proportional to the Kolmogorov--Sinai entropy (KS--entropy) $h_{KS}$
\begin{eqnarray}\label{0-0}
\tau_{KS}=\frac{1}{h_{KS}}
\end{eqnarray}
Another important property is the relationship between the maximum Lyapunov exponent and $h_{KS}$.
Krylov observed
that a little phase volume $\Delta V$ after a time $t$ will be
spread over a region with a volume $\Delta V(t)=\Delta V\exp(h_{KS}t)$ where
$\Delta V(t)$ is of order $1$ \cite{krylov1,krylov2}. This means that after a time
\begin{eqnarray}\label{0-1}
t_0=\frac{1}{h_{KS}}\ln \frac{1}{\Delta V}
\end{eqnarray}
the initial phase volume $\Delta V$ is spread over
the whole phase space. Consequently, one might expect that the typical
relaxation times are proportional to $\frac{1}{h_{KS}}$.

On the other hand, the Pesin theorem relates the KS-entropy $h_{KS}$ with the Lyapunov exponents by means of the formula \cite{lich,pesin,young,gutz}
\begin{eqnarray}\label{0-2}
h_{KS}=\int_{\Gamma}\sum_{\sigma_i>0}\sigma_i(q,p)dqdp
\end{eqnarray}
where $\Gamma$ is the phase space.
For the special case where the $\sigma_i$ are constant over all phase space one has
\begin{eqnarray}\label{0-3}
h_{KS}=\sum_{\sigma_i>0}\sigma_i
\end{eqnarray}
It should be noted the interest of the formula \eqref{0-2} and its physical meaning. Pesin theorem relates the KS-entropy,
that is the average unpredictability of information of all possible trajectories in
the phase space, with the exponential instability of motion. Then, the main
content of Pesin theorem is that $h_{KS} > 0$ is a sufficient condition for the chaotic
motion. Using Eqs. \eqref{0-0} and \eqref{0-3} one obtains the following relationship between $\tau_{KS}$ and the Lyapunov exponents
\begin{eqnarray}\label{0-4}
\frac{1}{\tau_{KS}}=\sum_{\sigma_i>0}\sigma_i
\end{eqnarray}
In the next sections we will use this formula in order to obtain a relationship between the Lyapunov exponents and the poles of a non Hermitian Hamiltonian.

\subsection{Wigner transformation and non Hermitian quantum dynamics}
We recall some properties of the Wigner transformation formalism \cite{Wigner gaussian,Wigner,Symb} and we give the notions of the non Hermitian quantum dynamics we will use throughout the paper. Given a quantum operator $\hat{A}$ the Wigner transformation $W_{\hat{A}}:\mathbb{R}^{2M}\mapsto\mathbb{R}$ of $\hat{A}$ is defined by
\begin{eqnarray}\label{wigner}
W_{\hat{A}}(q,p)=\frac{1}{h^{M}}\int_{\mathbb{R}^{M}}\langle q+\Delta|\,\hat{A}
\,|q-\Delta\rangle e^{2i\frac{p\Delta}{\hbar}}d\Delta
\end{eqnarray}
where $q,p,\Delta\in\mathbb{R}^M$.
The Weyl symbol $\widetilde{W}_{\hat{A}}:\mathbb{R}^{2M}\mapsto\mathbb{R}$ of $\hat{A}$ is defined by $\widetilde{W}_{\hat{A}}(q,p)=\hbar^{M}W_{\hat{B}}(q,p)$ where $\hbar=\frac{h}{2\pi}$ and $h$ is the Planck constant.
In particular, for the identity operator $\hat{I}$ one has
$\widetilde{W}_{\hat{I}}(q,p)=1(q,p)$ where $1(q,p)$ is the function that is constantly equal to $1$.
One of the main properties of the Wigner transformation
is the expression of integrals over the phase space in terms of trace of operators by means of \cite{Wigner}
\begin{eqnarray}\label{wigner2}
\textrm{Tr}(\hat{A}\hat{B})=\int_{\mathbb{R}^{2M}}W_{\hat{A}}(q,p)\widetilde{W}_{\hat{B}}(q,p)dqdp
\end{eqnarray}
valid for all pair of operators $\hat{A},\hat{B}$ where $\hat{A}\hat{B}$ denotes the product of
$\hat{A}$ and $\hat{B}$ and $\textrm{Tr}(\ldots)$ is the trace operation.
Using the definition of the Weyl symbol it can be shown the following result that relates the Weyl symbols of an operator and of the same but evolved at a time $t$. The proof can be found in the Appendix.
\begin{lemma}\label{lemaweyl}
Let $\widetilde{W}_{\hat{A}}(q,p)$ be the Weyl symbol of an operator $\hat{A}$. Then the Weyl symbol of $\hat{A}(-t)=\hat{U}_t^{\dag}\hat{A}\hat{U}_t$ is
$\widetilde{W}_{\hat{A}}(q(t),p(t))$ where $(q(t),p(t))=(T_tq,T_tp)$ and $T_t$ is the classical evolution given by Hamilton equations. For all $t\in\mathbb{R}$ one has
\begin{eqnarray}\label{wigner3}
\widetilde{W}_{\hat{U}_t^{\dag}\hat{A}\hat{U}_t}(q,p)=\widetilde{W}_{\hat{A}}(q(t),p(t))  \ \ \ \  \forall \ (q,p)\in \mathbb{R}^2
\end{eqnarray}
where $\hat{A}(-t)=\hat{U}_{-t}\hat{A}\hat{U}_{-t}^{\dag}$,
$\hat{U}_t=e^{-i\frac{\hat{H}}{\hbar}t}$ is the evolution operator, and $\hat{U}_t^{\dag}$ is the Hermitian conjugate of $\hat{U}_t$.
\end{lemma}


We consider a quantum system $S$ described by a non Hermitian Hamiltonian $\hat{H}$ having a discrete complex spectrum where $E_1=\hbar\omega_1+i\gamma_1,\ldots,E_N=\hbar\omega_N+i\gamma_N$ are the complex eigenvalues.
The eigenvalues $E_k = \hbar\omega_k + i\gamma_k$ contain the eigeneregies $\hbar\omega_k$, and the resonance widths $-\gamma_k>0$ that are interpreted as
proportional to the decay characteristic times of the system \cite{moiseyev}.
The non--Hermiticity of $\hat{H}$ implies the existence of two basis of eigenvectors called $\{\langle \widetilde{1}|, \langle \widetilde{2}|,\ldots,\langle \widetilde{N}|\}$ \emph{left eigenvectors} and $\{|1\rangle, |2\rangle,\ldots,|N\rangle\}$ \emph{right eigenvectors} satisfying the relations \cite{gilary}
\begin{eqnarray}\label{1}
\hat{H}|j\rangle=E_j |j\rangle \ \ , \ \ \ \ \langle \widetilde{j}|\hat{H}=\langle \widetilde{j}|E_j^{*} \ \ \ \ \ \ \ \ j=1,\ldots,N
\end{eqnarray}
and
\begin{eqnarray}\label{2}
&\langle \widetilde{j}|k\rangle=\delta_{jk} \,\,\ \ \ \  \forall \  j,k=1,\ldots,N \nonumber\\
&\sum_{j=1}^N |j\rangle \langle\widetilde{j}|=\hat{I}
\end{eqnarray}
where $E_j^{*}$ denotes the complex conjugate of $E_j$ for all $j=1,\ldots,N$. The formulas in \eqref{2}
correspond to the bi--orthogonality and completeness conditions, respectively.

\section{Lyapunov exponents in a non Hermitian quantum dynamics}
Let $Q$ a quantum system described by a non Hermitian Hamiltonian $\hat{H}$ having a discrete complex spectrum $E_1=\hbar\omega_1+i\gamma_1,\ldots,E_N=\hbar\omega_N+i\gamma_N$
and a phase space $\Gamma\subseteq\mathbb{R}^{2M}$. Consider the dynamical system description used in classical mechanics given by $(\Gamma,P(\Gamma),\mu,\{T_t\}_{t\in\mathbb{R}})$ where
$P(\Gamma)$ is $\sigma$--algebra of subsets of $\Gamma$, $\mu$ is the Lebesgue measure, and $T_t$ is the classical evolution\footnote{Typically, the one given by the Hamilton equations.} over the phase space.
Let us take a little volume $\Delta V$ which is the measure of some set $A\subset\Gamma$. That is,
\begin{eqnarray}\label{3}
\Delta V=\mu(A)=\int_{\Gamma}1_A(q,p)dqdp
\end{eqnarray}
where $\mu$ is the Euclidean measure of $\mathbb{R}^{2M}$ and $1_A(q,p)$ is the characteristic function of $A$.
Let $\hat{A}$ be the quantum operator such that $W_{\hat{A}}(q,p)=1_A(q,p)$. Since $\Gamma\subseteq\mathbb{R}^{2M}$ and $\widetilde{W}_{\hat{I}}(q,p)=1(q,p)$ then using the Wigner property \eqref{wigner2} one can recast \eqref{3} as
\begin{eqnarray}\label{4}
&\textrm{Tr}(\hat{A})=\int_{\mathbb{R}^{2M}}1_A(q,p)1(q,p)dqdp=\Delta V
\end{eqnarray}
In turn, from \eqref{3} and by Lemma 2.1. it follows that the volume $\Delta V$ at time $t$ is
\begin{eqnarray}\label{5}
&\Delta V(t)=\mu(T_tA)=\int_{\Gamma}1_{T_tA}(q,p)dqdp=\int_{\Gamma}1_{A}(T_{-t}q,T_{-t}p)dqdp\nonumber\\
&=\int_{\Gamma}1_{A}(q(-t),p(-t))dqdp=\int_{\mathbb{R}^{2M}}1_A(q(-t),p(-t))1(q,p)dqdp
\end{eqnarray}
Since $W_{\hat{A}}(q,p)=1_A(q,p)$ then by eq. \eqref{wigner3} one has
\begin{eqnarray}\label{6}
&1_A(q(-t),p(-t))=W_{\hat{A}}(q(-t),p(-t))=W_{\hat{A}(t)}(q,p)
\end{eqnarray}
By the Wigner property \eqref{wigner2} and \eqref{6} one can express \eqref{5} as
\begin{eqnarray}\label{7}
&\Delta V(t)=\textrm{Tr}(\hat{A}(t))=\textrm{Tr}(\hat{U}_{t}\hat{A}\hat{U}_{t}^{\dag})
\end{eqnarray}
Without loss of generality we assume that the motion over phase space is bounded. Physically, since the non--Hermiticity of the Hamiltonian implies an open system dynamics then after the KS--time the system occupies a volume $S$ in phase space and the spreading of any volume ceases.
If one consider $\Delta V=\frac{\hbar}{S}$ as the initial condition then $\Delta V(t)$ represents the portion of $S$ occupied by the system at time $t$. As we mentioned in section 2, at time $t=t_0$ the volume has spread throughout over its bounded region which is expressed mathematically as $\Delta V(t_0)=1$. Note also that $\Delta V$ is the inverse of the quassiclassical parameter $q=\frac{S}{\hbar}$ where $S$ is of the order of magnitude of the classical action.
If one expands $\hat{A}$ in the basis of the left--right eigenvectors
\begin{eqnarray}\label{8}
\hat{A}=\sum_{i,j=1}^N a_{ij}|i\rangle \langle\widetilde{j}|
\end{eqnarray}
Then by the condition $\textrm{Tr}(\hat{A})=\frac{\hbar}{S}$ one obtains
\begin{eqnarray}\label{9}
\frac{\hbar}{S}=\sum_{i=1}^N a_{ii}
\end{eqnarray}
For the sake of simplicity we assume that the diagonal elements of $\hat{A}$ are all the same, i.e. $a_{ii}=a_0$ for all $i=1,\ldots,N$. From \eqref{9}
it follows that
\begin{eqnarray}\label{10}
a_{ii}=\frac{1}{N}\frac{\hbar}{S} \ \ \ \ \ \  \forall \ i=1,\ldots,N
\end{eqnarray}
Then $\hat{A}$ at time $t$ is
\begin{eqnarray}\label{11}
&\hat{A}(t)=\hat{U}_t\hat{A}\hat{U}_t^{\dag}=\sum_{i,j=1}^N a_{ij}\exp\left(\left(-i(\omega_i-\omega_j)+\frac{\gamma_i+\gamma_j}{\hbar}\right)t\right)|i\rangle \langle\widetilde{j}|\nonumber
\end{eqnarray}
From Eqs. \eqref{2}, \eqref{9}, \eqref{10}, and \eqref{11} one has
\begin{eqnarray}\label{12}
\Delta V(t)=\frac{1}{N}\Delta V\sum_{i=1}^N\exp\left(2\frac{\gamma_i}{\hbar}t\right)
\end{eqnarray}
Now we are able to connect the Lyapunov exponents with the Hamiltonian poles $E_1=\hbar\omega_1+i\gamma_1,\ldots,E_N=\hbar\omega_N+i\gamma_N$.
By definition, if one sets $t=t_0$ in \eqref{12} it follows that
\begin{eqnarray}\label{13}
1=\frac{1}{N}\Delta V\sum_{i=1}^N\exp\left(2\frac{\gamma_i}{\hbar}t_0\right)
\end{eqnarray}
Therefore, using \eqref{0-1} and \eqref{0-2} in \eqref{13} one has
\begin{eqnarray}\label{14}
&\int_{\Gamma}\sum_{\sigma_i>0}\sigma_i(q,p)dqdp=\frac{1}{t_0}\log\left(\frac{1}{N}\sum_{i=1}^N\exp\left(2\frac{\gamma_i}{\hbar}t_0\right)\right)
\end{eqnarray}
which for the case $\sigma_i=\textrm{constant}$ for all $i$ becomes
\begin{eqnarray}\label{15}
\sum_{\sigma_i>0}\sigma_i=\frac{1}{t_0}\log\left(\frac{1}{N}\sum_{i=1}^N\exp\left(2\frac{\gamma_i}{\hbar}t_0\right)\right)
\end{eqnarray}
The equation \eqref{15} is the main result of the present contribution. It expresses the positive Lyapunov exponents $\sigma_i$ of the phase space dynamics in terms of the imaginary parts $\gamma_i$ of the Hamiltonian poles $E_i=\hbar\omega_i+i\gamma_i$.
It should be noted that since $\frac{\hbar}{S}<1$ then $\Delta V(t_0)=1$ is greater than the initial volume $\Delta V(0)=\frac{\hbar}{S}$. Thus, the $\gamma_i$ cannot be all negative.

\section{The model and results}
\subsection{The Gamow model}
In order to illustrate the physical relevance of the formula \eqref{15} we apply it to an example of the decoherence literature: a phenomenological Gamow model type \cite{PRE,Omnes1}. This model consists of a single oscillator embedded in an environment composed of a large bath of noninteracting oscillators, which can be considered as a continuum.
The degeneration of this system prevents the application of perturbation theory. Instead, we can apply an analytical extension of the Hamiltonian \cite{PRE,antoniou,Gadella,letterpolos,Ordonito-dec,4A} to obtain an non Hermitian effective Hamiltonian $\hat{H}_{eff}$
given by
\begin{eqnarray}\label{gamow3}
\hat{H}_{eff}=\sum_{n=0}^{\infty}z_{n}|n\rangle
\langle\widetilde{n}|
\end{eqnarray}
where $z_{n}=n(\hbar\omega_0-i\gamma_0)$ are complex eigenvalues, except $z_0=\omega_0$, and $\gamma_0>0$ is associated with the decoherence time $t_R=\frac{\hbar}{\gamma_0}$ and $\omega_0$ is the natural frequency of the single oscillator. The two set of eigenvectors $\{\langle \widetilde{m}|\}_{m=0}^{\infty}$ and $\{|n\rangle\}_{n=0}^{\infty}$ satisfy the bi-orthogonality and completeness relations given by \eqref{2}.
For numerical calculations we can always consider a truncated basis composed by $N+1$ eigenvectors
$|0\rangle,|1\rangle,\ldots,|N\rangle$ that simply bounds the motion of the single oscillator from zero energy up to a maximum value of energy equal to $\hbar(N+1)\omega_0$.

\subsection{Mapping contractions into expansions and viceversa by means of the time reversal system}
Since all the imaginary parts $\textrm{Im}(z_n)=-n\gamma_0$ are negative then by the last paragraph of previous section one can not apply the Eqs. \eqref{14} and \eqref{15} to obtain the Lyapunov coefficients. However, one can use the following strategy. The key is to consider an ``artificial" system $S^{\prime}$
which is the original but with the time evolution inverted, i.e. by performing the time transformation $t\rightarrow-t$. From here onwards, we will call ``time reversal system" to $S^{\prime}$.
Then, in order to apply the Eqs. \eqref{13}, \eqref{14}, and \eqref{15} on $S^{\prime}$ one simply should replace $2\frac{\gamma_i}{\hbar}t_0$ by
$2\frac{\gamma_i}{\hbar}(-t_0)=2\frac{-\gamma_i}{\hbar}t_0$. This simply means that in presence of complex eigenvalues the time reversal transformation $t\rightarrow-t$ is equivalent to change the sign of the imaginary parts of the eigenvalues, i.e. $\gamma_i\rightarrow-\gamma_i$.
Thus, in non Hermitian quantum mechanics the time invariance symmetry is satisfied only if one adds the transformation $\gamma_i\rightarrow-\gamma_i$.
Using the eq. \eqref{13} in $S^{\prime}$ one obtains
\begin{eqnarray}\label{gamow4}
&(N+1)=\Delta V\sum_{k=0}^N\exp\left(2\frac{\gamma_k}{\hbar}t_0\right) \nonumber\\
&\textrm{with} \ \    \gamma_k=k\gamma_0\geq0  \ \ \textrm{for all} \ \ k=0,\ldots,N
\end{eqnarray}
For solving \eqref{gamow4} is useful to adimensionalize $t_0$ by expressing it in terms of the relaxation time $t_R=\frac{\hbar}{\gamma_0}$. Taking into account this, the equation to be solved for $T_0$ is
\begin{eqnarray}\label{gamow5}
(N+1)=\Delta V\sum_{k=0}^N\exp(2kT_0) \ \ , \ \ T_0=\frac{t_0}{t_R}
\end{eqnarray}
where $T_0$ is the adimensionalized KS--time. Then, from Eqs. \eqref{0-1}, \eqref{0-3} and $T_0$ one can rewrite the KS--entropy in the convenient form
\begin{eqnarray}\label{gamow6}
h_{KS}=\frac{1}{T_0}\frac{\gamma_0}{\hbar}\log\frac{1}{\Delta V}=\sum_{\sigma_i^{\prime}>0}\sigma_i^{\prime}
\end{eqnarray}
It should be noted that $h_{KS}$ and the $\sigma_i^{\prime}$ are the KS--entropy and the Lyapunov exponents of the system $S^{\prime}$ respectively.
The Lyapunov exponents of the original system can be recovered by using the following argument. Since a positive (negative resp.) Lyapunov exponent $\sigma$ implies an expansion (contraction resp.) of some region of phase space then one has that $t\rightarrow-t$ maps $\sigma$ into $-\sigma$. From this it follows that $-\sigma_i^{\prime}$ are the Lyapunov exponents of the original system. In other words, the time reversal transformation $t\rightarrow-t$ maps contractions into expansions and viceversa.

\subsection{Adimensionalized KS--time of the time reversal system}
Now we solve numerically the eq. \eqref{gamow5} for a given number of oscillators $N$ and for some representative initial volumes $\Delta V$.
We analyze two cases: first we vary the number $N$ from $5$ to $100$ in steps of $\Delta N=5$. Second, we consider $N$ from $1000$ to $10000$ with $\Delta N=1000$. In both cases the chosen initial volumes $\Delta V$ are $10^{-3},10^{-6},10^{-9}$, and $10^{-12}$.
The first case is suitable to give an idea for the effects of using a finite basis while in the second case the situation is closer to a continuum bath of oscillators.

From the Table I one can see that $T_0$ decreases along with the number of oscillators $N$ and this is independent of the initial volume $\Delta V$, as expected. For a given $N$ the effect of $\Delta V$ is to increase the value of $T_0$ as soon as $\Delta V$ decreases. We can give an intuitive explanation about this. Since $T_0$ is the time that takes for $\Delta V$ to spread over the whole phase space then the more smaller is $\Delta V$, more bigger is $T_0$. Moreover, an interplay between $N$ and $\Delta V$ is observed. For instance, one can see that the same value of $T_0=0.0015$ is obtained for $N=3000$, $\Delta V=10^{-3}$ and for $N=10000$, $\Delta V=10^{-12}$. Physically, this means that any decrease in the initial volume can be compensated by an increase in the number of oscillators, i.e. if one wants to decrease $T_0$ then one must add more oscillators to the bath. The same situation is observed for $N=30$, $\Delta V=10^{-3}$ and for $N=100$, $\Delta V=10^{-12}$.

\begin{table}[ht]
\begin{center}
\begin{tabular}{|c|c|c|c|c|}
\hline
\multicolumn{5}{|c|}{adimensionalized KS--time $T_0[\frac{\hbar}{\gamma_0}]$} \\ \hline
$N$ & $10^{-3}$ & $10^{-6}$ & $10^{-9}$ & $10^{-12}$  \\
\hline \hline
5 & $0.85$ & $1.56$ & 0.0313 & 0.0313\\ \hline
10 & $0.438$ & $0.799$ & 1.15 &1.5\\ \hline
30 & $0.15$ & $0.287$ & 0.393 &0.511\\ \hline
60 & $0.0837$ & $0.146$ & 0.198 &0.257\\ \hline
100 & $0.0544$ & $0.0828$ & 0.119 &0.154\\ \hline
1000 & $0.0045$ & $0.0083$ & 0.0112 & 0.0155\\ \hline
3000 & $0.0015$ & $0.0027$ & 0.004 &0.0051\\ \hline
7000 & $0.0006$ & $0.0012$ & 0.0017 &0.0022\\ \hline
10000 & $0.0004$ & $0.0008$ & 0.0012 &0.0015\\ \hline
\end{tabular}
\end{center}
\caption{Some adimensionalized KS--times of the time reversal system in function of the number $N$ of the bath oscillators and for the initial volumes $\Delta V=10^{-3},10^{-6},10^{-9}$ and $10^{-12}$.}
\end{table}

\subsection{KS--entropy of the time reversal system}
Having computed numerically the adimensionalized KS--time of $S^{\prime}$ as a function of the number of the bath oscillators one can proceed to obtain the KS--entropy $h_{KS}$ of $S^{\prime}$. If one replaces the values of $T_0$ of the Table I in eq. \eqref{gamow6} then it results that $h_{KS}$ as a function of $N$ can be linearly adjusted for each value of the initial volume
\begin{eqnarray}\label{gamow8}
&h_{KS}(N)_{\Delta V=10^{-3}}= (1.5152\pm 0.0001) N \nonumber\\
&h_{KS}(N)_{\Delta V=10^{-6}}= (1.662 \pm 0.001) N \nonumber\\
&h_{KS}(N)_{\Delta V=10^{-9}}= (1.7355 \pm 0.001) N  \nonumber\\
&h_{KS}(N)_{\Delta V=10^{-12}}= (1.778\pm 0.001) N
\end{eqnarray}
From eq. \eqref{gamow8} it is straightforward that one can deduce the approximated formula
\begin{eqnarray}\label{gamow9}
h_{KS}(N)= (1.5\pm 0.3) N  \ \ \ \ \  (\textrm{in units of} \ \frac{\gamma_0}{\hbar})
\end{eqnarray}
for the KS--entropy of the time reversal system $S^{\prime}$ which is valid for all the range in $N$ and $\Delta V$ studied.


\subsection{Lyapunov exponents in terms of Hamiltonian poles}
Now we can proceed to obtain the Lyapunov exponents of the Gamow model in term of its Hamiltonian poles. From
eqs. \eqref{gamow6} and \eqref{gamow9} it follows that
\begin{eqnarray}\label{gamow10}
(1.5\pm 0.3) N \frac{\gamma_0}{\hbar}=\sum_{\sigma_i^{\prime}>0}\sigma_i^{\prime}>0
\end{eqnarray}
where $\sigma_i^{\prime}$ are the Lyapunov exponents of the time reversal system $S^{\prime}$. The physical meaning of \eqref{gamow10} is straightforwardly to explain. The adimensional characteristic time $T_0$ is inversely proportional to $N$ and since $h_{KS}$ is inversely proportional to $T_0$ then $h_{KS}$ is a linear and increasing function of the number of bath oscillators. This implies that the effect of each oscillator of the bath is to increase $h_{KS}$ in an amount of $(1.5\pm 0.3)\frac{\gamma_0}{\hbar}$ where $N=0$ corresponds to the single oscillator $\omega_0$ without the presence of the bath oscillators.
Moreover, since the oscillators of the bath are non interacting then the effect of all the bath oscillators is simply the sum of each of them. It follows that each oscillator of the bath contributes with a same Lyapunov exponent, namely $\sigma_0^{\prime}$, in such way that the sum of right hand in \eqref{gamow10} becomes
\begin{eqnarray}\label{gamow11}
\sum_{\sigma_i^{\prime}>0}\sigma_i^{\prime}>0=N\sigma_0^{\prime}
\end{eqnarray}
From Eqs. \eqref{gamow10} and \eqref{gamow11} one obtains
\begin{eqnarray}\label{gamow12}
\sigma_0^{\prime}=(1.5\pm 0.3) \frac{\gamma_0}{\hbar}
\end{eqnarray}
This is the Lyapunov exponent of the time reversal system $S^{\prime}$, which is positive since all the volumes $\Delta V=\frac{\hbar}{S}$ expand along their allowed region of phase space after the KS--time $T_0$. In particular, the transformation $t\rightarrow-t$ maps $\sigma_0^{\prime}$ into $-\sigma_0^{\prime}$. Now we can arrive to the other main result of this paper. The Lyapunov exponent $\sigma_0$ of the Gamow model in terms of its Hamiltonian poles is given by
\begin{eqnarray}\label{gamow13}
\sigma_0=-\alpha \frac{\gamma_0}{\hbar}    \ \ \ \ \ \ \ \ \ \ \alpha=(1.5\pm 0.3)
\end{eqnarray}
which is negative according to its dissipative behavior. The factor $\alpha$ can be interpreted as a coupling constant which is characteristic of the bath. Alternatively, the Lyapunov exponent also can be expressed in terms of the relaxation time $t_R=\frac{\hbar}{\gamma_0}$ as
\begin{eqnarray}\label{gamow14}
\sigma_0=-\alpha \frac{1}{t_R}    \ \ \ \ \ \ \ \ \ \ \alpha=(1.5\pm 0.3)
\end{eqnarray}

\subsection{Limit cases}

From the Eqs. \eqref{gamow13} and \eqref{gamow14} one can analyze two limit cases. The first one results by considering the limit $\gamma_0\rightarrow0$ that simply corresponds to
a single harmonic oscillator of frequency $\omega_0$. It is clear that in such case there is no dissipation, without expansion nor contraction of volumes in the phase space. Therefore, all the Lyapunov exponents must be zero. This is precisely what it is obtained by setting $\gamma_0=0$ in \eqref{gamow13}, i.e. $\sigma_0=0$.
The other limit case results by considering that the relaxation time is vanishingly small. This corresponds to a maximal dissipation where the oscillator is fully damped by the bath. In such a case the Lyapunov exponent is infinitely negative, as one obtains by taking the limit
$t_R\rightarrow0$ in the formula \eqref{gamow14}.

\section{A discussion at the light of the quantum resonances theory}

Here we provide a discussion of the connection between some approaches in quantum resonances theory \cite{semi2,semi3,semi4} and the framework presented in this paper. Several previous work based on the formalism of the resonance gap lower bounds \cite{res4,res5,res6}, semiclassical
approaches based on short periodic orbits in open systems \cite{res7}, and phenomenology by means of a
mixture of phase-space dynamics \cite{res9} among others, show that a unified theory of resonance still appears to be a difficult task.

Nevertheless, we can mention the aspects of our proposal in agreement with some standard approaches used in the description of quantum open systems. Below we quote some results of the literature and discuss them from the point of view of the present paper.

\begin{itemize}
  \item \emph{Semiclassical periodic-orbit theory \cite{semi2}: Studies about quantum
scattering resonances of dissociating molecules have reported that signatures of the classical bifurcation appear in the spectrum of resonances. These can be obtained as generalized eigenstates, whose eigenergies are denoted by $E_n=\varepsilon_n-i\Gamma_n/2$, of a non Hermitian Hamiltonian operator, as in Eqs. \eqref{1} and \eqref{2}. In turn, the Gutzwiller trace allows to give a semiclassical description of the resonances as the complex zeros of
\begin{eqnarray}\label{5-1}
S(E)=2\pi\hbar(n+\frac{\mu}{4})-i\frac{\hbar}{2}T(E)\Lambda(E)
\end{eqnarray}
with $n=0,1,2\ldots$, $\mu=2$ for the symmetric--stretch period orbit, $S(E)$ denotes the reduced action $\oint \textbf{p}.d\textbf{q}$,
and $\Lambda(E)$ stands for Lyapunov exponent in the vicinity of the periodic orbit of period $T(E)$. Since the imaginary part $\Gamma_n/2$ is usually smaller than the real part $\varepsilon_n$ then one of the successes of this approach is that, in \eqref{5-1}, one can expand the action $S(E)$ around the energy of the resonances as $S(\varepsilon_n)=2\pi\hbar(n+1/2)$. In such a way that the widths of the resonance are determined
by the Lyapunov exponent $\lambda(\varepsilon_n)$ of the periodic orbit
\begin{eqnarray}\label{5-2}
\lambda(\varepsilon_n)=\frac{\Gamma_n}{\hbar}=\frac{1}{\tau_n}
\end{eqnarray}
where $\tau_n$ are the lifetimes of the quantum resonances.}

Looking at the eqs. \eqref{gamow13}, \eqref{gamow14}, this is precisely what we have obtained for the case of the Gamow model that has only a single relevant lifetime given by its decoherence time $t_R=-\frac{\alpha}{\sigma_0}$. Moreover, with the help of Eqs. \eqref{10}, \eqref{11}, \eqref{12}, \eqref{13}, \eqref{14} one can also recover the formula \eqref{5-2}. Due to Eqs. \eqref{10} and \eqref{11}, if the single oscillator is at the $n$--th level then it has an energy $n\hbar\omega_0$, which corresponds to a superposition of $n$ bath oscillators, and the Lyapunov exponent $\lambda_n$ for the unstable $n$--orbit results
\begin{eqnarray}\label{5-3}
\lambda_n=-n\alpha\frac{\gamma_0}{\hbar}=\frac{1}{t_n} \ \ \ \ \ , \ \ \ \ t_N=\frac{t_R}{n}
\end{eqnarray}
where $t_n$ is the lifetime corresponding to the quasi--stationary state $|N\rangle$ for all $n=1,2,3,\ldots$

\item \emph{Generalized Pesin theorem \cite{semi3}: in the context of chaotic open systems the Pesin formula \eqref{0-3} can be generalized as
\begin{eqnarray}\label{5-4}
H_{KS}=\sum_{\sigma_i>0}\sigma_i - \gamma=h_{KS}-\gamma
\end{eqnarray}
where $\gamma$ is the escape rate of the trajectories leaving the system and $H_{KS}$ denotes the KS--entropy that takes into account $\gamma$.}

One can see that the time reversal technique used in section 4 is in agreement with the generalized Pesin formula \eqref{5-4}.
Since the lifetime of a trajectory corresponding to the $N$--th energy level is proportional to $\gamma_N/\hbar=N\gamma_0/\hbar$ (with the proportionality factor given by the coupling constant $\alpha$ and $\gamma_N$ the $N$--resonance width) then the escape rate $\gamma$ is $\alpha N\gamma_0/\hbar$. Thus, the generalized Pesin theorem implies that
\begin{eqnarray}\label{5-5}
H_{KS}=\sum_{\sigma_i^{\prime}>0}\sigma_i^{\prime}-\alpha N\gamma_0/\hbar=0
\end{eqnarray}
which is nothing but the eq. \eqref{gamow10}.

\item \emph{Classical localization of chaotic resonances states \cite{semi4}: in order to describe the classical localization of chaotic states of quantum systems, a conditionally invariant measure $\mu_{\gamma}$ is defined by
\begin{eqnarray}\label{5-6}
\mu_{\gamma}(T^{-1}A)=e^{-\gamma}\mu_{\gamma}(A)
\end{eqnarray}
for all subset $A$ of phase space. }

Taking into account the developed in section 3, i.e. the way of expressing volumes in phase space as traces of quantum operators, the measure $\mu_{\gamma}$ can be obtained with an explicit expression for $\gamma$ in terms of the imaginary parts $\gamma_i$. From Eqs. \eqref{3}, \eqref{5}, and \eqref{12} and for $t=1$ it follows that

\begin{eqnarray}\label{5-7}
\mu_{\gamma}(T(A))=e^{-\gamma}\mu(A)   \ \ \ \ \ , \ \ \ \ \gamma=\gamma(\gamma_1,\ldots,\gamma_N)=-\log\left(  \frac{1}{N}\sum_{i=1}^N\exp\left(2\frac{\gamma_i}{\hbar}t\right)  \right)
\end{eqnarray}
Now if one applies the time reversal transformation (or, equivalently, by changing the sign of the imaginary parts $\gamma_i$) then the transformation $T$ must be replaced by $T^{-1}$, and therefore, \eqref{5-7} becomes \eqref{5-6}.

\end{itemize}

\section{Conclusions}
We have presented a relationship between the Lyapunov exponents and the Hamiltonian poles in a non Hermitian dynamics. We have deduced this relationship, the eq. \eqref{15}, by means of the Pesin theorem and the KS--time, and with the help of expressing volumes in phase space as traces of quantum operators. We have illustrated the formalism with a phenomenological Gamow model type and the results have been interpreted and linked with those obtained by using other approaches in the literature.
The relevance of our contribution lies in several aspects, which we enumerate below:
\begin{itemize}
  \item \emph{Resonances and decoherence}: the characteristic decay times given by the imaginary part of the complex eigenvalues can be connected with the Lyapunov exponents concerning the dynamics in phase space, in agreement with the semiclassical periodic--orbit theory \cite{semi2} but using a simpler mathematics. Moreover, for the Gamow model the decoherence time is inversely proportional to the Lyapunov exponents of the unstable periodic orbits (eq. \eqref{5-3}).
  \item \emph{KS--entropy in non Hermitian Hamiltonian systems}: one has a method to obtain the part of the KS--entropy, free of the escape rates, of a quantum system having a non Hermitian Hamiltonian. For the Gamow model this results equivalent to using the generalized Pesin theorem \cite{semi3}.


  \item \emph{Lyapunov exponents of dissipative systems}: the use of the time reversal system could provide an indirect way to obtain the negative Lyapunov exponents of a dissipative system, as was accomplished for the Gamow model.
\item \emph{Invariant measure for classical localization}: the use of the Wigner function to express classical quantities as quantum traces allows to give the \emph{conditionally invariant measure} (CIM, \cite{semi4}) in terms of the decay modes of the quasi stationary states, i.e. as a function of the imaginary part of the complex eigenvalues.
\end{itemize}

We hope the results of this work can be useful to shed light on the search for a unified theory of quantum resonances.

\section*{Acknowledgments}
This work was partially supported by CONICET and Universidad Nacional de La Plata, Argentina.

\appendix
\section{Proof of Lemma 2.1.}
\begin{proof}
By definition, one has
\begin{eqnarray}\label{prop1}
\widetilde{W}_{\hat{A}}(q,p)=\int_{\mathbb{R}}\langle q+\Delta| \hat{A}|q-\Delta\rangle e^{2i\frac{p\Delta}{\hbar}}d\Delta
\end{eqnarray}
Then it follows that
\begin{eqnarray}\label{prop2}
\widetilde{W}_{\hat{A}}(T_{\epsilon }q,T_{\epsilon }p)=\int_{\mathbb{R}}\langle T_{\epsilon }q+\Delta| \hat{A}|T_{\epsilon }q-\Delta\rangle e^{2i\frac{T_{\epsilon }p\Delta}{\hbar}}d\Delta
\end{eqnarray}
where $T_{\epsilon}$ is a the transformation $T_t$ at time $t=\epsilon$. if one consider $|\epsilon|\ll1$ then $T_{\epsilon}$ is approximately equal to the identity function of the phase space $\Gamma$, i.e. $T_{\epsilon}\approx T_0=1_{\Gamma}$.
Now if one make the change of variables $\widetilde{\Delta}=T_{-\epsilon}\Delta$, then
\begin{eqnarray}\label{prop3}
\Delta=T_{\epsilon}\widetilde{\Delta}  \ \ \ \ \textrm{and} \ \ \ \ \
d\Delta=|T_{\epsilon}|d\widetilde{\Delta}
\end{eqnarray}
where $|T_{\epsilon}|$ is the Jacobian determinant of $T_{\epsilon}$ restricted to the coordinates $q$.
Using \eqref{prop3} one can recast \eqref{prop2} as
\begin{eqnarray}\label{prop4}
\widetilde{W}_{\hat{A}}(T_{\epsilon}q,T_{\epsilon}p)=\int_{\mathbb{R}}\langle T_{\epsilon}q+T_{\epsilon}\widetilde{\Delta}| \hat{A}|T_{\epsilon}q-T_{\epsilon}\widetilde{\Delta}\rangle e^{2i\frac{T_{\epsilon}pT_{\epsilon}\widetilde{\Delta}}{\hbar}}|T_{\epsilon}|d\widetilde{\Delta}
\end{eqnarray}
It is clear that
\begin{eqnarray}\label{prop5}
\langle T_{\epsilon}q+T_{\epsilon}\widetilde{\Delta}|=\langle q+\widetilde{\Delta}|\hat{U}^{\dag}(\epsilon) \ \ \ \ \textrm{and} \ \ \ \ \
|T_{\epsilon}q-T_{\epsilon}\widetilde{\Delta}\rangle=\hat{U}(\epsilon)|q-\widetilde{\Delta}\rangle
\end{eqnarray}
Also,
\begin{eqnarray}\label{prop6}
&e^{2i\frac{T_{\epsilon}pT_{\epsilon}\widetilde{\Delta}}{\hbar}}=e^{2i\frac{p\widetilde{\Delta}}{\hbar}}\Longleftrightarrow
\frac{T_{\epsilon}pT_{\epsilon}\widetilde{\Delta}}{\hbar}-\frac{p\widetilde{\Delta}}{\hbar}=m\pi \ \
\Longleftrightarrow \ \ \ p(\epsilon)\widetilde{\Delta}(\epsilon)-p\widetilde{\Delta}=mh/2 \nonumber\\
&\textrm{with} \ \ \ m\in\mathbb{Z} \ \ \
\textrm{and} \ \ \ p(\epsilon)=T_{\epsilon}p \ \ , \ \ \widetilde{\Delta}(\epsilon)=T_{\epsilon}\widetilde{\Delta}
\end{eqnarray}
By considering the Planck constant $h$ vanishingly small then
the eq. \eqref{prop6} is satisfied.
Thus, if one replaces \eqref{prop5} and \eqref{prop6} in \eqref{prop4} it follows that
\begin{eqnarray}\label{prop7}
&\widetilde{W}_{\hat{A}}(T_{\epsilon}q,T_{\epsilon}p)=\int_{\mathbb{R}}\langle q+\widetilde{\Delta}|\hat{U}^{\dag}(\epsilon) \ \hat{A} \ \hat{U}(\epsilon)|q-\widetilde{\Delta}\rangle e^{2i\frac{p\widetilde{\Delta}}{\hbar}}|T_{\epsilon}|d\widetilde{\Delta}
\end{eqnarray}
Moreover, if one applies the change of variables theorem to the variables $\Delta,\widetilde{\Delta}$ then one can express \eqref{prop7} as
\begin{eqnarray}\label{prop8}
\widetilde{W}_{\hat{A}}(T_{\epsilon}q,T_{\epsilon}p)=\int_{\mathbb{R}}\langle q+\Delta|\hat{U}^{\dag}(\epsilon) \ \hat{A} \ \hat{U}(\epsilon)|q-\Delta\rangle e^{2i\frac{p\Delta}{\hbar}}d\Delta=\widetilde{W}_{\hat{U}^{\dag}(\epsilon)\hat{A} \hat{U}(\epsilon)}(q,p)=\widetilde{W}_{\hat{A}(-\epsilon)}(q,p)
\end{eqnarray}
valid for all $\epsilon$ arbitrarily small.
Also, for an arbitrary $t\in\mathbb{R}$ one has
\begin{eqnarray}\label{prop9}
T_t=T_{\epsilon}^N         \ \ \ \text{with} \ \ \ t=N\epsilon
\end{eqnarray}
where $T_{\epsilon}^N$ denotes the composition of $T_{\epsilon}$ with itself $N$ times.
Then, by iterating the formula \eqref{prop8} $N$ times and using \eqref{prop9} the desired result is obtained.

\end{proof}

\end{document}